\documentclass[english,showpacs,floatfix]{revtex4}
\usepackage{amsfonts}
\usepackage{amssymb,epsf}
\usepackage{latexsym}

\usepackage[dvips]{graphicx}
\usepackage{longtable}
\usepackage{amsmath,amssymb}
\usepackage{dcolumn}
\usepackage{latexsym}
\usepackage{babel}
\usepackage[latin1]{inputenc}

\newcommand{\be}{\begin{equation}}
\newcommand{\ee}{\end{equation}}
\newcommand{\no}{\nonumber}

\begin{document}

\title{String inspired explanation for the super-acceleration of our universe}

\author{Ahmad Sheykhi $^{1,2,3}$\footnote{
asheykhi@mail.uk.ac.ir} and  Bin Wang $^{1}$\footnote{
wangb@fudan.edu.cn}}

\address{$^1$  Department of Physics, Fudan University, Shanghai 200433, China\\
         $^2$  Department of Physics, Shahid Bahonar University, Kerman, Iran}
 \author{Nematollah Riazi \footnote{
riazi@physics.susc.ac.ir}}
\address{$^3$Physics Department and Biruni Observatory, Shiraz University,
Shiraz 71454, Iran}

\begin{abstract}
We investigate the effect of the bulk content in the general
Gauss-Bonnet braneworld on the evolution of the universe. We find
that the Gauss-Bonnet term and the combination of the dark
radiation and the matter content of the bulk play a crucial role
in the universe evolution. We show that our model can describe the
super-acceleration of our universe with the equation of state of
the effective dark energy in agreement with observations.

\end{abstract}


\maketitle
\section{Introduction}\label{Intr}
One of the most dramatic discoveries of the modern cosmology in
the past decade is that our universe is currently accelerating
\cite{Rie}. A component that causes the accelerated expansion of
the universe is referred to as ``dark energy". Although the nature
of such dark energy is still speculative, an overwhelming flood of
papers has appeared which attempt to describe it by devising a
great variety of models. Among them are cosmological constant,
exotic fields such as phantom or quintessence, modified gravity,
etc, see \cite{Pad}\cite{Cop} for a recent review. Available
models of dark energy differ in the value and variation of the
equation of state parameter $w$ during the evolution of the
universe. The cosmological constant with $w=-1$, is located at a
central position among dark energy models both in theoretical
investigation and in data analysis \cite{Wei}. In quintessence
\cite{QUINT}, Chaplygin gas \cite{KMP} and holographic dark energy
models \cite{MLi}, $w$ always stays bigger than $-1$. The phantom
models of dark energy have $w<-1$ \cite{PHANT}. However, following
the more accurate data analysis, a more dramatic result appears
showing that the time varying dark energy gives a better fit than
a cosmological constant and in particular, $w$ can cross $-1$
around $z=0.2$ from above to below \cite{Alam}. Although the
galaxy cluster gas mass fraction data do not support the
time-varying $w$ \cite{chen}, theoretical attempts towards the
understanding of the $w$ crossing $-1$ phenomenon have been
started. Some dark energy models, such as the one containing a
negative kinetic scalar field and a normal scalar field
\cite{Feng}, or a single scalar field model \cite{MZ} and
interacting holographic dark energy models \cite{Wang1} have been
constructed to gain insight into the occurrence of the transition
of the dark energy equation of state and the mechanism behind this
transition. Other studies on the $w=-1$ crossing have been carried
out in \cite{Noj}.

Independent of the challenge we deal with the dark energy puzzle,
in recent years, theories of large extra dimensions, in which the
observed universe is realized as a brane embedded in a higher
dimensional spacetime, have received a lot of interest. According
to the braneworld scenario the standard model of particle fields
are confined on the brane while, in contrast, the gravity is free
to propagate in the whole spacetime. In these theories the
cosmological evolution on the brane is described by an effective
Friedmann equation that incorporates non-trivially with the
effects of the bulk onto the brane. An interesting consequence of
the braneworld scenario is that it allows the presence of
five-dimensional matter which can propagate in the bulk space and
may interact with the matter content in the braneworld. It has
been shown that such an interaction can alter the profile of the
cosmic expansion and lead to a behavior that would resemble the
dark energy. The cosmic evolution of the braneworld with energy
exchange between brane and bulk has been studied in different
setups \cite{Kirit}\cite{Kof}\cite{Cai}\cite{Bog}. In these
models, due to the energy exchange between the bulk and the brane,
the usual energy conservation law on the brane is broken and
consequently it was found that the equation of state of the
effective dark energy may experience the transition behavior (see
e.g \cite{Cai,Bog}).

On the other hand, in string theory, in addition to the Einstein
action, some higher derivative curvature terms have been included
to derive the gravity. In order to obtain a ghost-free theory, the
combination of quadratic terms called Gauss-Bonnet term is usually
employed as curvature corrections to the Einstein-Hilbert action
\cite{Zwi}. From a geometric point of view, the combination of the
Einstein-Hilbert and Gauss-Bonnet term constitutes, for 5D
spacetimes, the most general Lagrangian to produce second-order
field equations \cite{Lov}. The Gauss-Bonnet correction
significantly changes the bulk field equations and leads to
modifications in the braneworld Friedmann equations. Therefore,
the study of the effects of the Gauss-Bonnet correction term on
the evolution of the universe in the braneworld scenario is well
motivated. Influences of the Gauss-Bonnet correction on the DGP
braneworld have been studied in \cite{maartens, cai2}.

The purpose of the present work is to investigate the effects of
the bulk content in the general Gauss-Bonnet braneworld on the
evolution of the universe. Although the effects of the
Gauss-Bonnet correction term on the late time universe is small,
we will see that it still plays an important role in the cosmic
evolution. Besides we will show that the combination of the dark
radiation term and the matter content of the bulk plays the role
of the dark energy on the brane and influences the evolution of
the universe. In our model, in contrast to the previous models
(\cite{Kirit,Kof,Cai,Bog}), we do not need to break down the
standard energy momentum conservation law on the brane, although
our model can allow such assumption if one is interested. We will
show that by suitably choosing model parameters, our model can
exhibit accelerated expansion of the universe. In addition, we
will present a profile of the $w$ crossing $-1$ phenomenon which
is in good agreement with observations.

The paper is organized as follows. In Section \ref{model}, we
present a braneworld model to describe the accelerated expansion
and the effective equation of state of dark energy in the presence
of the Gauss-Bonnet correction term in the bulk. In Section
\ref{late}, we study the cosmological consequences of the model
and in particular, its effect on the evolution of the universe.
The last section is devoted to conclusions and discussions.

\section{The model}\label{model}
The theory we are considering is five-dimensional and has an
action of the form
\begin{eqnarray}\label{Act}
S  =  \frac{1}{2{\kappa}^2} \int{
d^5x\sqrt{-{g}}\left({R}-2\Lambda+\alpha \mathcal{L}_{GB}\right)}
+\int{ d^5x\sqrt{-{g}}\mathcal{L}_{bulk}^{m}}+\int
{d^{4}x\sqrt{-\tilde{g}}(\mathcal {L}_{brane}^{m}-\sigma)},
\end{eqnarray}
where $\Lambda<0$ is the bulk cosmological constant and $\mathcal{
L}_{GB}$ is the Gauss-Bonnet correction term
\begin{equation}
\mathcal{L}_{GB} = R^2-4R^{AB}R_{AB}+R^{ABCD}R_{ABCD}\,.
\end{equation}
Here $g$ and $\tilde{g}$ are the bulk and brane metrics,
respectively. $R$, $R_{AB}$, and $R_{ABCD}$ are the scalar
curvature and Ricci and Riemann tensors, respectively. Throughout
this paper we choose the unit so that $\kappa^2=1$ as the
gravitational constant in five dimension. We have also included
arbitrary matter content both in the bulk and on the brane through
$\mathcal{L}_{bulk}^{m}$ and $\mathcal {L}_{brane}^{m}$
respectively, and $\sigma$ is the positive brane tension. The
field equations can be obtained by varying the action (\ref{Act})
with respect to the bulk metric $g_{AB}$. The result is
\begin{eqnarray}
G_{AB}+\Lambda g_{AB}+2\alpha H_{AB} = T_{AB}, \label{Feq}
\end{eqnarray}
where $H_{AB}$ is the second-order Lovelock tensor
\begin{eqnarray}
H_{AB} & = & RR_{AB}-2R_A{}^CR_{BC}-2R^{CD}R_{ACBD} \nonumber \\
& & +R_A{}^{CDE}R_{BCDE}-\textstyle{1\over4}g_{AB}{\cal L}_{GB}
\,. \nonumber
\end{eqnarray}
For convenience and without loss of generality, we can choose the
extra-dimensional coordinate $y$ such that the brane is located at
$y=0$ and bulk has $\mathbb{Z}_2$ symmetry. We are interested in
the cosmological solution with a metric
\begin{eqnarray}
ds^2&=&-n^2(t,y) dt^2 + a^2(t,y)\gamma_{ij}dx^i dx^j +
b^2(t,y)dy^2, \label{line}
\end{eqnarray}
where $\gamma _{ij}$ is a maximally symmetric $3$-dimensional
metric for the surface ($t$=const., $y$=const.), whose spatial
curvature is parameterized by k = -1, 0, 1. The metric
coefficients $n$ and $b$ are chosen so that, $n(t,0)=1$ and
$b(t,0)=1$, where $t$ is cosmic time on the brane. The total
energy-momentum tensor has bulk and brane components and can be
written as
\begin{equation}
{T}_{AB}=
{T}_{AB}\mid_{brane}+{T}_{AB}\mid_{\sigma}+{T}_{AB}\mid_{bulk}.
\end{equation}
The first and the second terms are the contribution from the
energy-momentum tensor of the matter field confined to the brane
and the brane tension
\begin{eqnarray}
T^{A}_{\,\,B}\mid_{brane}\,&=&\,\mathrm{diag}(-\rho,p,p,p,0)\frac{\delta(y)}{b},{\label{bem}}\\
T^{A}_{\,\,B}\mid_{\sigma}\,&=&\,\mathrm{diag}(-\sigma,-\sigma,-\sigma,-\sigma,0)\frac{\delta(y)}{b},{\label{sigma}}
\end{eqnarray}
where $\rho$, and $p$, being the energy density and pressure on
the brane, respectively. In addition we assume an energy-momentum
tensor for the bulk content of the form
\begin{equation}
T^{A}_{\ B}\mid_{bulk}\,= \,\left(\begin{array}{ccc}
T^{0}_{\ 0}\,&\,0\,&\,T^{0}_{\ 5}\\
\,0\,&\,T^{i}_{\ j}\delta^i_{\ j}\,&\,0\\
-\frac{n^2}{b^2}T^{0}_{\ 5}\,&\,0\,&\,T^{5}_{\ 5}
\end{array}\right)\,\,.\,\,\,
\end{equation}
The quantities which are of interest here are $T^{5}_{\ 5}$ and
$T^{0}_{\ 5}$, as these two enter the cosmological equations of
motion. In fact, $T^{0}_{\ 5}$ is the term responsible for energy
exchange between the brane and the bulk.  Integrating the $(00)$
component of the field equations (\ref{Feq}) across the brane and
imposing $\mathbb{Z}_2$ symmetry, we have the jump across the
brane \cite{kofin}
\begin{eqnarray}\label{Jc1}
\left[ 1+4\alpha\left(H^2+\frac{k}{a_{0}^{2} }- \frac{a^{\prime
\,2}_{+}}{3a_{0}^2} \right)\right]\frac{a'_{+}}{ a_{0} }
 =-\frac{1}{6}(\rho+\sigma),
\end{eqnarray}
where $ 2a'_{+}=-2a'_{-}$ is the discontinuity of the first
derivative. $H=\dot{a}_{0}/a_{0}$ is the Hubble parameter on the
brane. Eq. (\ref{Jc1}) is a cubic equation for the discontinuity
$a'_{+}/a_{0}$, which has only one real solution, the other two
being complex. Therefore, if we require our cosmological equations
to have the right $\alpha\rightarrow 0$ limit we are left with
only one solution. However, this real root is too lengthy and
complicated to present here. Since we are interested to study the
effect of the Gauss-Bonnet correction term on the evolution of the
universe in the late time so it is reasonable to choose the
Gauss-Bonnet coupling constant $\alpha$ to be small, namely
$0<\alpha<1$. Using this fact we can expand the real solution for
$a'_{+}/a_{0}$ versus $\alpha$ powers. The result for $k=0$ up to
order $\alpha$ is
\begin{eqnarray}\label{ju1}
\frac{a'_{+}}{a_{0}}&=&-\frac{1}{6}(\rho+\sigma)+{\frac
{\alpha}{162}}\,\, \left( \rho+\sigma
 \right)  \left( 108\,{H}^{2}- \left( \rho+\sigma \right) ^{2}
 \right)+O(\alpha^2).
\end{eqnarray}
In a similar way, integrating the $(ij)$ component of the field
equations (\ref{Feq}) across the brane and imposing $\mathbb{Z}_2$
symmetry, we can obtain the discontinuity in the metric function
$n'_{+}/n_{0}$, which for $k=0$ can be written up to O($\alpha$)
in the following form
\begin{eqnarray}\label{ju2}
\frac{n'_{+}}{n_{0}} &=&\frac{1}{6}(2\rho+3p-\sigma)+\frac
{\alpha}{3} \left( -2\,{H}^{2}
 \left( 2\,\rho+3\,p-\sigma \right)\right. \no
\\
&& \left. +{\frac {1}{54}}\, \left( \rho+ \sigma \right) ^{2}
\left( 8\,\rho+9\,p-\sigma \right) +4\,\dot{H} \left( \rho+\sigma
\right) \right)+O(\alpha^2),
\end{eqnarray}
where dots denote time derivatives and primes denote derivatives
with respect to $y$. At this point we find it convenient to absorb
the brane tension $\sigma$ in $\rho$ and $p$ with the replacement
$\rho+\sigma\rightarrow \rho$ and $p-\sigma\rightarrow p$.
Therefore the junction conditions $(\ref{ju1})$ and $(\ref{ju2})$
can be simplified
\begin{eqnarray}\label{jun3}
\frac{a'_{+}}{a_{0}}&=&-\frac{\rho}{6}+{\frac {\alpha}{162}}\,\,
 \rho\left( 108\,{H}^{2}- \rho^{2}\right),  \\
 \frac{n'_{+}}{n_{0}} &=&\frac{1}{6}(2\rho+3p)+\frac
{\alpha}{3} \left( -2\,{H}^{2}
 \left( 2\,\rho+3\,p\right)+{\frac {\rho^{2}}{54}}\,
\left( 8\,\rho+9\,p\right)+4\,\dot{H} \rho\right).\label{jun4}
\end{eqnarray}
Substituting the junction conditions $(\ref{jun3})$ and
$(\ref{jun4})$ into the $(55)$ and $(05)$ components of the field
equations $(\ref{Feq})$, we obtain the modified Friedmann equation
and the semi-conservation law on the brane (up to order $\alpha$)
\begin{eqnarray}\label{fri1}
&&H^2\left(1-\frac{\alpha}{9}\rho\left(2\rho+3p \right)\right)+\no
\left(\dot{H}+H^2\right)\left(1+4\alpha
\left(H^2+\frac{\rho^2}{36}\right)\right) \no\\
&&+\frac{\rho}{36}\left(\rho+3p\right)
+\frac{\alpha}{972}\rho^3\left(2\rho+3p\right)=\frac{\Lambda-T^{5}_{\
5}}{3},
\end{eqnarray}
and
\begin{eqnarray}\label{T1}
\dot{\rho}+3H(\rho+p)=-T,  \   \    \    \ T\equiv 2 T^{0}_{\
5}\left[1-4\alpha\left(H^2-\frac{\rho^2}{36}\right)\right].
\end{eqnarray}
We shall assume an equation of state $p=w\rho$ to hold between the
energy density and pressure of matter on the brane. Therefore we
have
\begin{eqnarray}\label{fri2}
&&H^2\left(1-\frac{\alpha}{9} \rho^2\left(2+3\omega
\right)\right)+\left(\dot{H}+H^2\right)\left(1+4\alpha
\left(H^2+\frac{\rho^2}{36}\right)\right)\no \\
&&+\frac{\rho^2}{36}\left(1+3\omega\right)
+\frac{\alpha}{972}\rho^4\left(2+3\omega\right)=\frac{\Lambda-T^5_{\ 5}}{3},\\
&&\dot{\rho}+3H\rho(1+\omega)=-T. \label{T2}
\end{eqnarray}
One can easily check that in the limit $\alpha\rightarrow0$, Eqs.
(\ref{ju1})-(\ref{T2}) reduce to the corresponding equations of
the braneworld model without Gauss-Bonnet correction term
\cite{Kirit}.

Remarkably, we can show that the Friedmann equation (\ref{fri2})
is equivalent to the following equations
\begin{eqnarray}\label{fri3}
2\alpha H ^{4}+ \left(1+\frac{\alpha \rho^{2}}{9}\right)
{H}^{2}=\frac{{\rho}^{2}}{36}\left(1+ \frac{\alpha
\rho^{2}}{54}\right)+\chi+\frac{\Lambda}{6}-\frac{T^5_{\ 5}}{3},
\end{eqnarray}
with $\chi$ satisfying
\begin{eqnarray}
\dot {\chi}+ 4\,H\left(\chi-\frac{T^5_{\
5}}{6}\right)=\frac{2}{36}\,{T}\,\rho\,
 \left[ 1-4\,\alpha\, \left( {H}^{2}-{\frac {{\rho}^{2}}{108}}\,
 \right)  \right]+\frac{\dot{T}^5_{\ 5}}{3}.
\end{eqnarray}
Using the definition for $T$ in Eq. (\ref{T1}), the latter
equation up to order $\alpha$ can be written as
\begin{eqnarray}\label{chi}
\dot {\chi}+ 4\,H\left(\chi-\frac{T^5_{\
5}}{6}\right)=\frac{4}{36}\,{T^0_{5}}\,\rho\,
 \left[ 1-8\,\alpha\, \left( {H}^{2}-{\frac {{\rho}^{2}}{54}}\,
 \right)  \right]+\frac{\dot{T}^5_{\ 5}}{3}.
\end{eqnarray}
Eq. (\ref{fri3}) is the modified Friedmann equation describing
cosmological evolution on the brane. The auxiliary field $\chi$
incorporates non-trivial contributions of dark energy which differ
from the standard matter fields confined to the brane. The bulk
matter contributes to the energy content of the brane through the
bulk pressure terms $T^{5}_{\ 5}$ that appear in the right hand
side of the Friedmann equation. In addition, the bulk matter
contributes to the energy conservation equation (\ref{T1}) through
$T^{0}_{ \ 5}$ which is responsible for the energy exchange
between the brane and bulk. The functions $T^{5}_{\ 5}$ and
$T^{0}_ {\ 5}$ are functions of time corresponding to their values
on the brane. The energy-momentum conservation
 $\nabla_AT^A_{\,\,B}=0$ cannot fully determine $T^{5}_{\ 5}$ and
$T^{0}_ {\ 5}$ and a particular model of the bulk matter is
required \cite{Bog}. In the limit $\alpha\rightarrow 0$, Eqs.
(\ref{fri3}) and (\ref{chi}) reduce to ( after replacement $\rho
\rightarrow\rho+\sigma$)
\begin{eqnarray}
{H}^{2}=\frac{(\rho+\sigma)^{2}}{36}+\chi+\frac{\Lambda}{6}-\frac{T^5_{\ 5}}{3}, \\
\dot {\chi}+ 4\,H\left(\chi-\frac{T^5_{\
5}}{6}\right)=\frac{4}{36}\,{\it
T^0_{5}}\,(\rho+\sigma)+\frac{\dot{T}^5_{\ 5}}{3}\,.
 \end{eqnarray}
If we invoke the usual definitation $\beta\equiv1/{36}$,
$\lambda\equiv(\Lambda+{\sigma^2}/{6})/6$ and
$\gamma\equiv\sigma\beta$, we get ($\kappa^2=1$)
 \begin{eqnarray}\label{RS}
{H}^{2}=\beta\rho^2+2\gamma\rho+\lambda+\chi-\frac{T^5_{\ 5}}{3}, \no\\
\dot {\chi}+ 4\,H\left(\chi-\frac{T^5_{\
5}}{6}\right)=4T^0_{5}(\beta\rho+\gamma)+\frac{\dot{T}^5_{\
5}}{3},
 \end{eqnarray}
which is noting, but the general set of the equations in RS II
braneworld model with bulk matter content plus brane-bulk energy
exchange (see for example \cite{Bog}).

Returning to the general Friedman equation (\ref{fri3}) with
Gauss-Bonnet correction term, we can show that this equation has
the solution for $H$ of the form
\begin{eqnarray}
H^2=-\frac{1}{4\alpha}-\frac{\rho^2}{36}\pm{\frac
{1}{108\alpha}}\,\left[ 729+12\alpha \rho^{2}(27+\alpha
\rho^{2})+972\alpha(6\chi+\Lambda-2 T^{5}_{\ 5})\right]^{1/2}.
 \end{eqnarray}
The upper solution (+) has correct $\alpha\rightarrow 0$ limit.
Indeed, if we expand this solution verses $\alpha$ we get (up to
O($\alpha$))
\begin{eqnarray}\label{fri4}
{H}^{2}=\frac{\rho^{2}}{36}+\chi +\frac{\Lambda}{6}-\frac{T^{5}_{\
5}}{3}-\frac{\alpha}{18}
\left[2\rho^{2}\left(\frac{\rho^{2}}{27}+2\chi+\frac{\Lambda}{3}-\frac{2T^{5}_{\
5}}{3}\right)+\left(6{\chi}+\Lambda-2T^{5}_{\ 5}\right)^2\right],
\end{eqnarray}
and Eqs. (\ref{T1}) and (\ref{chi}) become
\begin{eqnarray}
&&\dot{\rho}+3H\rho(1+\omega)=-2T^0_{\
5}\left[1-4\alpha\left(\chi+\frac{\Lambda}{6}-\frac{T^{5}_{\
5}}{3}\right)\right],\label{T3}\\
&&\dot {\chi}+ 4\,H\left(\chi-\frac{T^5_{\
5}}{6}\right)=\frac{4}{36}\,{T^0_{\ 5}}\,\rho\,
 \left[ 1-8\,\alpha\, \left( {\frac {{\rho}^{2}}{108}}+\frac{\Lambda}{6}-\frac{T^{5}_{\
5}}{3}+\chi
 \right)  \right]+\frac{\dot{T}^{5}_{\
5}}{3}.\label{chi2}
\end{eqnarray}
Therefore, until now we have obtained  the set of equations
describing the dynamics of our universe (Eqs.
(\ref{fri4})-(\ref{chi2})) in the general Gauss-Bonnet braneworld
with both bulk matter content and bulk-brane energy exchange
provided that the Gauss-Bonnet coupling constant $\alpha$ is
chosen sufficiently small. It is worth noting that although
$\alpha$ is small, it has a dramatic effect on the dynamic
behavior of the cosmic evolution. Besides the appearance of the
$\rho^4$ term on the right hand side of Eq. (\ref{fri4}) shows
that in high energy scale the Gauss-Bonnet correction term plays
an important role.

\begin{figure}[tbp]
\includegraphics[width=13cm]{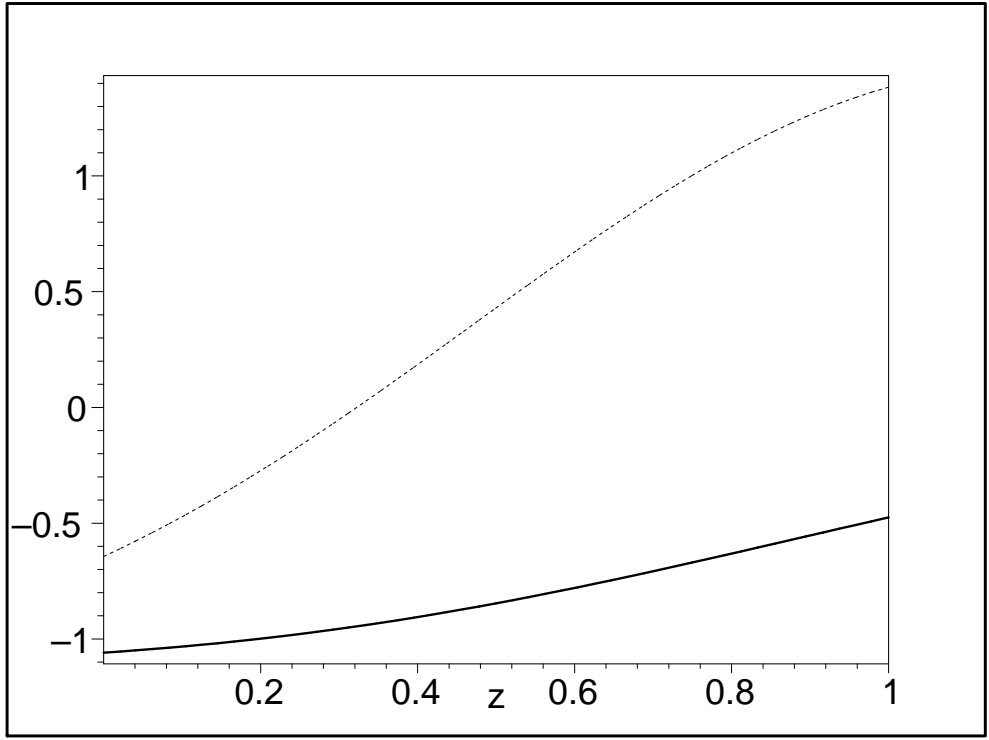}\\
 \caption{Evolution of
$w_{\mathrm{eff}}(z)$ (bold line)  and  $q(z)$ (dashed line)
versus $z$ for $\nu=0.34$ and $\protect\alpha=0$.}
\label{FigureRS}
\end{figure}
\begin{figure}[tbp]
\includegraphics[width=13cm]{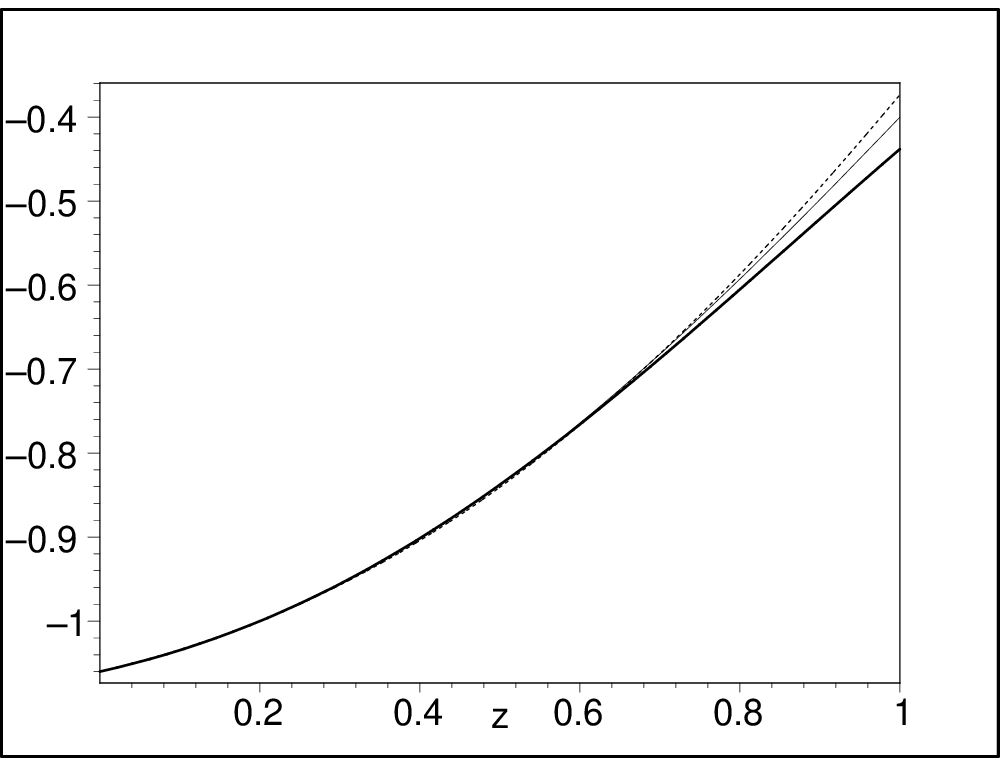}\\
\caption{Evolution of $w_{\mathrm{eff}}(z)$  versus  $z$ for
$0<\nu\leq0.17$. $\protect\alpha=0.01$ (bold line),
$\protect\alpha=0.1 $ (continuous line), and $\protect\alpha=0.9$
(dashed line).} \label{Figure1}
\end{figure}
\begin{figure}[tbp]
\includegraphics[width=13cm]{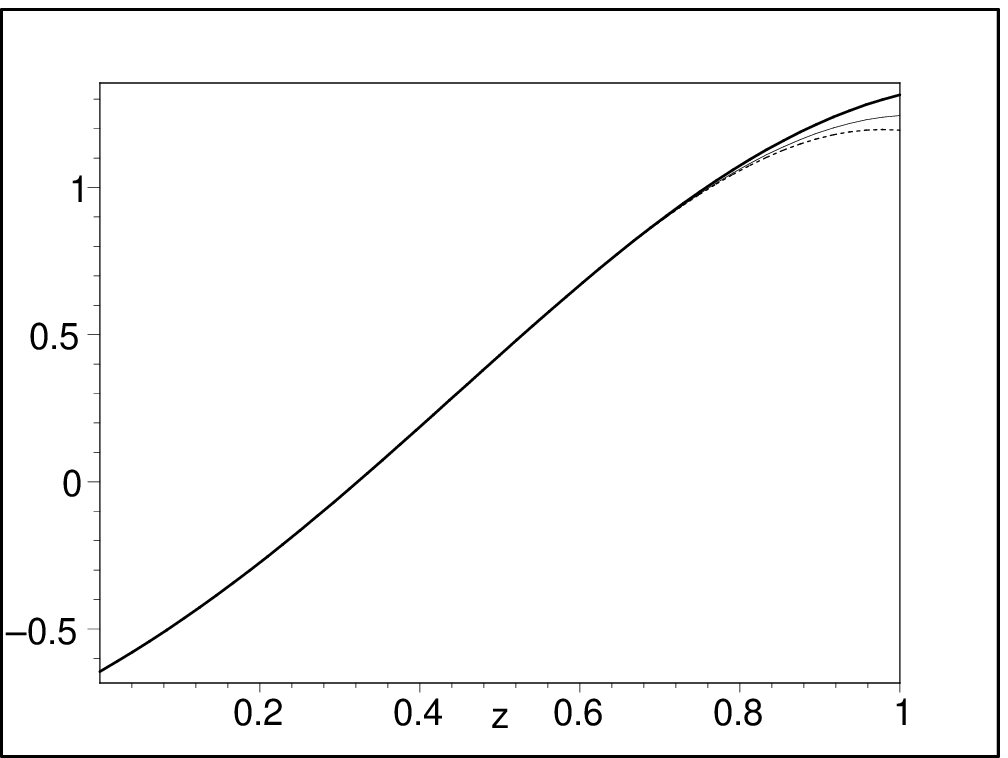}\\
\caption{Evolution of $q(z)$  versus  $z$ for $0 < \nu \leq 0.17$.
$\protect\alpha=0.01$ (bold line), $\protect\alpha=0.1 $
(continuous line), and $\protect\alpha=0.9$ (dashed line).}
\label{Figure2}
\end{figure}
\begin{figure}[tbp]
\includegraphics[width=13cm]{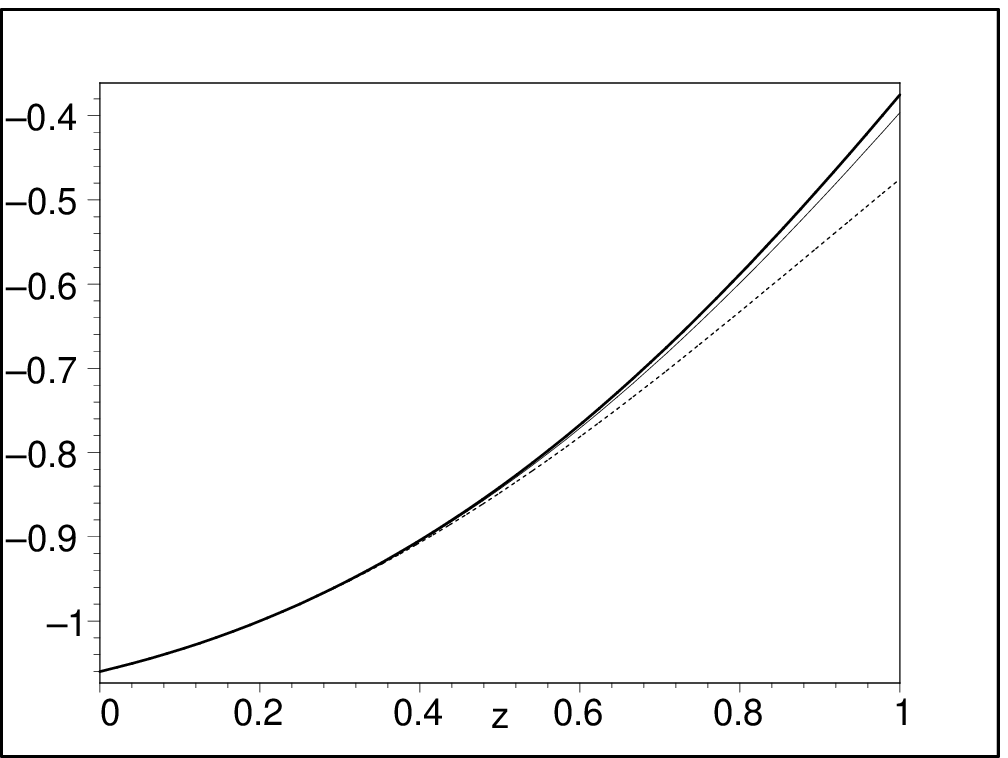}\\
\caption{Evolution of $w_{\mathrm{eff}}(z)$  versus  $z$ for
$0.18\leq\nu\leq0.34$. $\protect\alpha=0.01$ (bold line),
$\protect\alpha=0.1 $ (continuous line), and $\protect\alpha=0.9$
(dashed lined).} \label{Figure3}
\end{figure}
\begin{figure}[tbp]
\includegraphics[width=13cm]{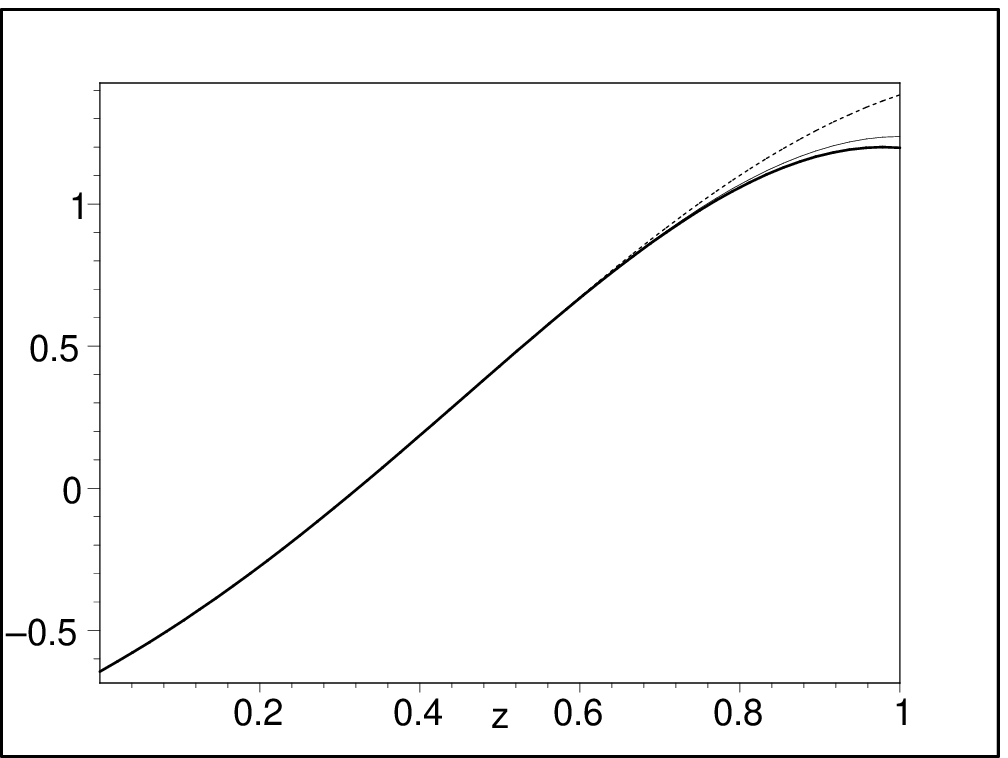}\\
\caption{Evolution of $q(z)$   versus  $z$ for
$0.18\leq\nu\leq0.34$. $\protect\alpha=0.01$ (bold line),
$\protect\alpha=0.1 $ (continuous line), and $\protect\alpha=0.9$
(dashed line).} \label{Figure4}
\end{figure}\begin{figure}[tbp]
\includegraphics[width=13cm]{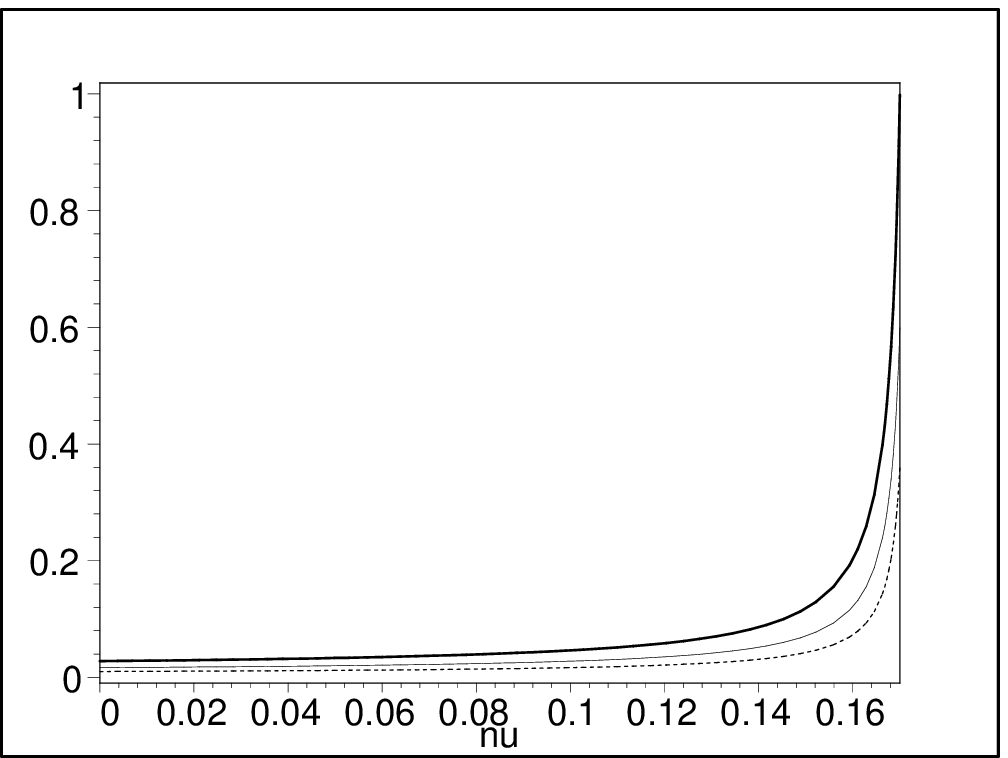}\\
\caption{ The parameter space of the function
$\protect\alpha(\nu,A)$ for $0 < \nu \leq 0.17$. $A=18$ (bold
line), $A=30 $ (continuous line), and $A=50$ (dashed line).}
\label{Figure5}
\end{figure}
\begin{figure}[tbp]
\includegraphics[width=13cm]{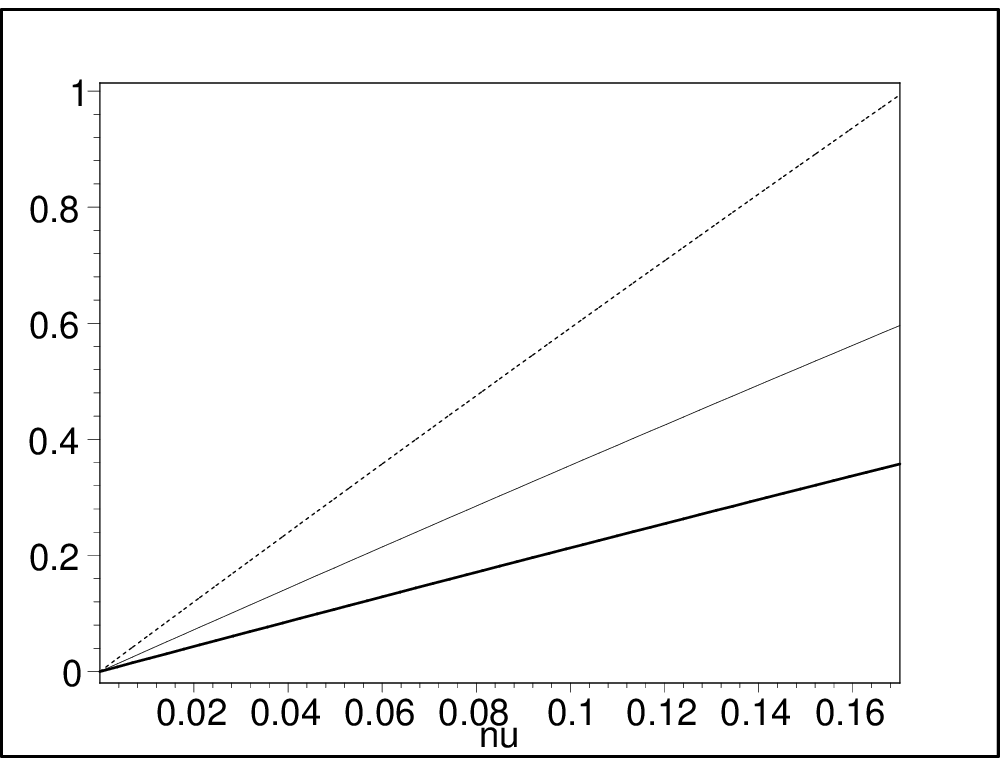}\\
\caption{ The parameter space of the function $C(\nu,A)$ for $0 <
\nu \leq 0.17$. $A=18$ (bold line), $A=30 $ (continuous line), and
$A=50$ (dashed line).} \label{Figure6}
\end{figure}
\begin{figure}[tbp]
\includegraphics[width=13cm]{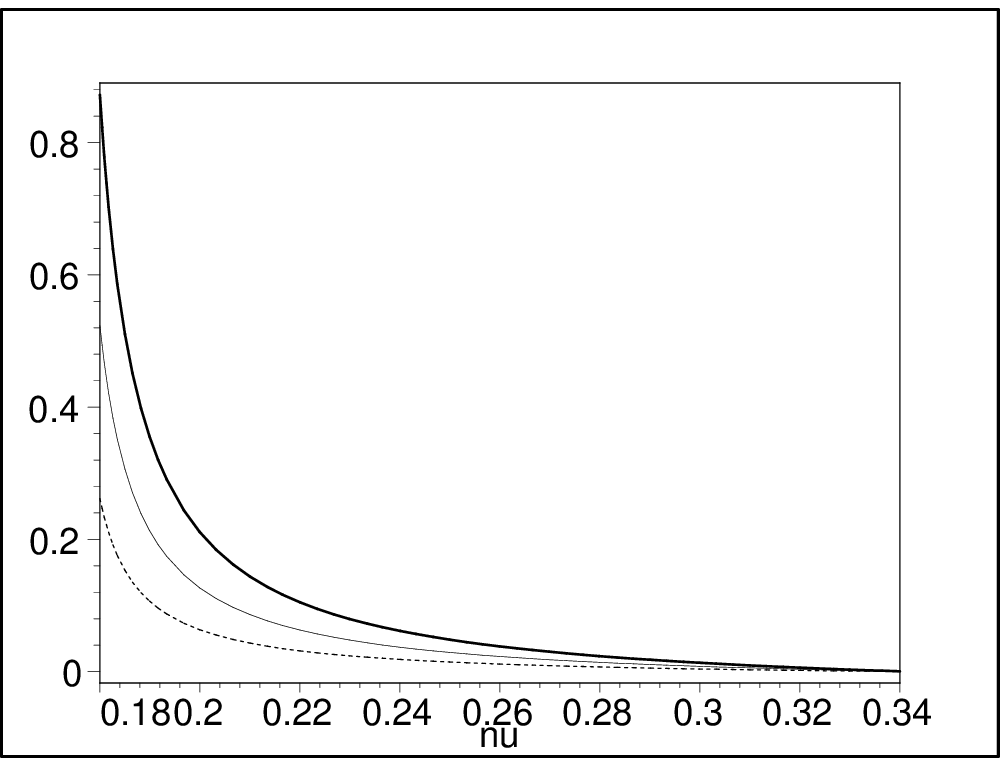}\\
\caption{ The parameter space of the function
$\protect\alpha(\nu,A)$ for $0.18 \leq \nu \leq 0.34$. $A=-6$
(bold line), $A=-10 $ (continuous line), and $A=-20$ (dashed
line). } \label{Figure7}
\end{figure}
\begin{figure}[tbp]
\includegraphics[width=13cm]{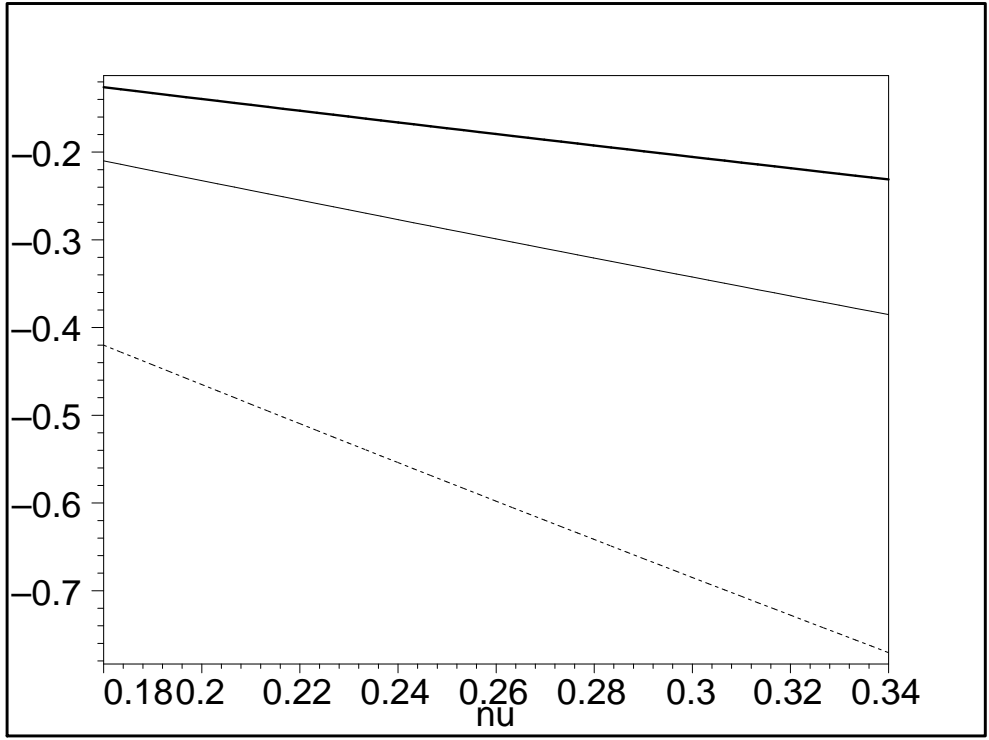}\\
\caption{ The parameter space of the function $C(\nu,A)$ for $0.18
\leq \nu \leq 0.34$. $A=-6$ (bold line), $A=-10 $ (continuous
line), and $A=-20$ (dashed line).} \label{Figure8}
\end{figure}

\section{COSMOLOGICAL CONSEQUENCES}\label{late}
In this section we are going to explore some cosmological
consequences of our model. To do this, first we separate back the
matter energy density and the brane tension as usual form with the
replacement $\rho \rightarrow\rho+\sigma$. Therefore Eqs.
(\ref{fri4}) and (\ref{chi2}) become
\begin{eqnarray}\label{Fri5}
{H}^{2}&=&\frac{2\sigma
\rho}{36}\left(1+\frac{\rho}{2\sigma}\right)+\frac{1}{6}\left(\Lambda+\frac{\sigma^{2}}{6}\right)-\frac{T^{5}_{\
5}}{3}+\chi
\nonumber\\
&&-\frac{\alpha}{18}
\left[2\sigma^{2}\left(1+\frac{\rho}{\sigma}\right)^2\left(\frac{\sigma^{2}}{27}\left(1+\frac{\rho}{\sigma}\right)^2+2\chi+\frac{\Lambda}{3}-\frac{2T^{5}_{\
5}}{3}\right) +\left(6{\chi}+\Lambda-2T^{5}_{\ 5}\right)^2\right],
\end{eqnarray}
\begin{eqnarray}\label{chi3}
 \dot {\chi}+ 4\,H\left(\chi-\frac{T^5_{\
5}}{6}\right)&=&\frac{4\sigma}{36}\,{T^0_{5}}\,\left(1+\frac{\rho}{\sigma}\right)\,
 \left[ 1-8\,\alpha\, \left( {\frac {{\sigma}^{2}}{108}}(1+\frac{\rho}{\sigma})^2+\frac{\Lambda}{6}-\frac{T^{5}_{\
5}}{3}+\chi
 \right)  \right]+\frac{\dot{T}^{5}_{\
5}}{3}.
\end{eqnarray}
We are interested in the scenarios where the energy density of the
brane is much lower than the brane tension, namely
$\rho\ll\sigma$. Assuming the Randall-Sundrum fine-tuning
$\Lambda+\sigma^2/6=0$  holds on the brane and defining the
parameter $\gamma\equiv{\sigma}/{36}$, Eqs. (\ref{Fri5}) and
(\ref{chi3}) can be simplified in the following form
\begin{eqnarray}\label{Fri6}
{H}^{2}&=& 2\gamma \rho+\chi-\frac{T^5_{\ 5}}{3}
-\frac{\alpha}{18} \left[2\left(\chi-\frac{T^5_{\
5}}{3}\right)\left(\sigma^2 + 18\left(\chi-\frac{T^5_{\
5}}{3}\right)\right)- \frac{\sigma^4}{108}\right],
\end{eqnarray}
\begin{eqnarray}\label{chi4}
 \dot {\chi}+ 4\,H\left(\chi-\frac{T^5_{\
5}}{6}\right)&=&4 \gamma \ {T^0_{\
5}}\,\left[1-8\alpha\left(\chi-\frac{T^5_{\
5}}{3}-\frac{\sigma^2}{54}\right)\right]+\frac{\dot{T}^{5}_{\
5}}{3}.
\end{eqnarray}
Now, one may adopt several strategies to find solutions of Eqs.
(\ref{T3}), (\ref{Fri6}) and (\ref{chi4}). For example, one may
take a suitable ansatz for the time dependent functions
${T}^{0}_{\ 5}$ and ${T}^{5}_{\ 5}$ and using Eq. (\ref{chi4}) to
find the function $\chi$. Then substitute $\chi$, ${T}^{0}_{\ 5}$
and ${T}^{5}_{\ 5}$ into Eq. (\ref{T3}) one can try to obtain
$\rho$, and finally one may find Hubble parameter $H$ through Eq.
(\ref{Fri6}). In the following we are interested in the case in
which the energy momentum conservation law on the brane holds,
which is usually assumed in the braneworld scenarios. Indeed, we
want to consider the effect of the bulk content on the evolution
of the universe without brane-bulk energy exchange, therefore we
set ${T}^{0}_{\ 5}=0$. The case with brane-bulk energy exchange in
the general Gauss-Bonnet braneworld will be addressed elsewhere.
It was argued that the energy exchange between the bulk and brane
${T}^{0}_{\ 5}$ will lead to the effective dark energy equation of
state crossing $-1$ [16,17]. Here we will show that without the
energy exchange, the effect of $T_5^5$ and the combined $T_5^5$
and the Gauss-Bonnet correction have the same role.

Inserting the condition ${T}^{0}_{\ 5}=0$ in Eq. (\ref{T3}), it
reduces to $\dot{\rho}+3H\rho(1+\omega)=0$. This equation has well
known solution $\rho=\rho_{0} a^{-3(1+w)}$, where $\rho_{0}$ is
the present matter density of the universe and we have omitted the
``o" subscript from the scale factor on the  brane for simplicity.
Then, consider a general ansatz ${T}^{5}_{\ 5}= D a^{\nu}$ for the
bulk pressure \cite{Bog}, where $D$ and $\nu$ are two arbitrary
constants, one can easily check that  Eq. (\ref{chi4}) has a
solution of the form
\begin{equation}\label{Chi}
\chi=C a^{-4}+Ba^{\nu},
\end{equation}
where $C$ is a constant usually referred to as dark radiation term
and $B\equiv D(\nu+2)/(3\nu+12)$. Finally, inserting $\rho$ and
$\chi$ into Eq. (\ref{Fri6}), we can rewrite it in the standard
form
\begin{equation}
H^2\,=\,\frac{8\pi
G_N}{3}(\rho+\rho_{\mathrm{eff}})\,,{\label{Fried}}
\end{equation}
where $G_N=3\gamma/4\pi$ is the $4$-dimensional Newtonian
constant and $\rho_{\mathrm{eff}}$ represents the effective dark
energy density on the brane
\begin{equation}{\label{rhoeff}}
\rho_{\mathrm{eff}}=\frac{1}{2\gamma}\left(C
a^{-4}+Aa^{\nu}\right)-\frac{\alpha}{36\gamma}\left[ 2\left(C
a^{-4}+Aa^{\nu}\right)\left({\sigma}^{2}+18\left(C a^{-4}+Aa^{\nu}
\right) \right)-\frac{{\sigma}^{4}}{108}\right],
\end{equation}
where $A\equiv-2D/(3\nu+12)$. The equation of state parameter of
the effective dark energy on the brane can be defined by
\cite{Lin}
\begin{equation}\label{eqw}
w_{\mathrm{eff}}=-1-\frac{1}{3}\frac{d\ln \delta H^2}{d\ln a},
\end{equation}
where $\delta H^2=(H^2/H_0^2)-\Omega_m a^{-3}$ accounts for terms
in the Friedmann equation except the brane matter with equation of
state $w_{m}=0$. Now, if we use the redshift parameter
$1+z=a^{-1}$ as our variable, we can easily show that
\begin{eqnarray}\label{w}
\omega_{\mathrm{eff}}(z)&=&-1+\frac{1}{3}\left(4\,C \left( 1+z
\right) ^{4}-A\nu\, \left( 1+z
 \right) ^{-\nu} \right)  \left[ 1-\frac{\alpha}{9}\, \left( 36\,A \left( 1
+z \right) ^{-\nu}+36\,C \left( 1+z \right) ^{4}+{\sigma}^{2}
\right)
 \right]  \no \\
 &&\times \Bigg{\{} C \left( 1+z \right) ^{4}+A \left( 1+z \right) ^{-\nu
} -\frac{\alpha}{18}\,\left[2\, \left( A \left( 1+z \right)
^{-\nu}+C
 \left( 1+z \right) ^{4} \right) \right. \no
\\
&& \left. \times \left( {\sigma}^{2}+18\,A \left( 1+z
 \right) ^{-\nu}+18\,C \left( 1+z \right) ^{4} \right) -{\frac {{\sigma}^{4}}{108
}}\, \right]\Bigg{\}}^{-1}.
\end{eqnarray}
The corresponding late time deceleration parameter can be written
\begin{equation}\label{q}
q (z)\equiv - \frac{1}{{H^2 }}\frac{{\ddot a}}{a}
=\frac{1}{2}\left[\Omega_{m}+(1-\Omega_m) \left(
1+3\omega_{\mathrm{eff}}(z)\right)\right],
\end{equation}
where $\Omega_{m}=\Omega_{m0}\, \left( 1+z \right) ^{3}$ is all
part of the matter on the braneworld and we take its present value
as $\Omega_{m0}=0.28\pm0.02$. In the rest of the paper, we will
obtain constraints on the parameters such as $C$, $A$, $\nu$,
$\alpha$ and $\sigma$ in our model. Indeed, we want to show that
under what parameter space constraints our model can describe the
accelerated expansion of the universe with the equation of state
of the effective dark energy $\omega_{\mathrm{eff}}$ crossing
$-1$, as suggested by observations.

\subsection{Special case with $\alpha=0$ }

Let us begin with the special case, in which the Gauss-Bonnet
coupling constant $\alpha$ is equal to zero. In this case we have
the usual Randall-Sundrum II  braneworld model and Eq. (\ref{w})
reduces to
\begin{equation}
w_{\mathrm{eff}}(z)=-1+\frac{1}{3}\,\left({\frac {4\,C \left( 1+z
\right) ^{\nu+4}-A\nu}{C \left( 1+z
 \right) ^{\nu+4}+A}}\right).
\end{equation}
Therefore, we are left with three parameters $C$, $A$, $\nu$, and
two of them are independent. Requiring that at the present moment
$w_{\mathrm{eff}}(z=0)=-1.06$ and $w$ crossed $-1$ around $z=0.2$
as indicated by extensive analysis of observational data
\cite{Alam}, we can obtain
\begin{equation}
{C}= 0.039 {A}, \hspace{0.5cm}   \nu=0.34, \hspace{0.5cm} A= A.
\end{equation}
For these value of parameters and $\Omega_{m0}=0.28$, from Eq.
(\ref{q}) we have $q(z=0)=-0.64$ and in addition $q(z)$ crosses
$0$ around $z=0.33$ which is in good agreement with recent
observational data \cite{Rie}\cite{aa}. In figure \ref{FigureRS}
we plot $w_{\mathrm{eff}}(z)$ and $q(z)$ for the above value of
the parameters versus redshift parameter $z$.

\subsection{General case with $\alpha\neq0$ }

Next, we consider the general Gauss-Bonnet braneworld with bulk
matter content. In this case we have five parameters only four of
which are independent. Considering that the value of $\sigma$ does
not affect the general profile of our model and further according
to the Randall-Sundrum fine-tuning relation it should be small, we
first fix $\sigma=10^{-3}$.  Thus, we have now four parameters and
among them three are independent. Numerical calculations show that
the functions $w_{\mathrm{eff}}$ and $q$ are well behaved for
$z\geq0$, provided that $0 < \nu \leq 0.34$. Employing the present
value of the equation of state parameter of dark energy
$\omega_{\mathrm{eff}}(z=0)=-1.06$ and the moment it crossed $-1$,
namely $\omega_{\mathrm{eff}}(z=0.2)=-1$, we get
\begin{equation}
 \alpha=\alpha(A,\nu), \hspace{0.6cm}C=0.12\, A  \nu\,({ 1.2})^{- \,\nu},  \hspace{0.6cm}
 A=A.
\end{equation}
If we impose the condition $0<\alpha<1$ which was used in deriving
our equations, we can get constraint on the free parameter $A$. In
numerical calculations we find that for $0<\nu\leq0.17$ we should
have $A>17.97$, while for $0.18\leq\nu\leq0.34$ we should have
$A<-5.23$ to satisfy the condition on $\alpha$. In figures
\ref{Figure1} and \ref{Figure2} we plot $w_{\mathrm{eff}}(z)$ and
$q(z)$ for $0<\nu\leq0.17$ versus redshift parameter $z$ for
different value of the Gauss-Bonnet coupling constant $\alpha$.
From these figures we observe that at large $z$, the
$w_{\mathrm{eff}}(z)$ increases with the increase of $\alpha$,
while $q(z)$ decreases with the increase of $\alpha$. This
qualitative behavior is quite opposite when $0.18\leq\nu\leq0.34$
as one can see from figures \ref{Figure3} and \ref{Figure4}.
Finally we plot in figures \ref{Figure5}-\ref{Figure8} the
parameter space for the functions $\alpha=\alpha(A,\nu)$ and
$C=C(A,\nu)$. We find that in the case $0<\nu\leq0.17$, $\alpha$
and $C$ increase with the increase of $\nu$ while, in contrast,
for $0.18\leq\nu\leq0.34$, $\alpha$ and $C$ decrease with the
increase of $\nu$.

\section{Conclusions and Discussions}\label{Conc}
In this work we have generalized the Randall-Sundrum II braneworld
with both bulk matter content and bulk-brane energy exchange by
adding the Gauss-Bonnet curvature correction term in the bulk
action. We have investigated the effects of the bulk content in
the general Gauss-Bonnet braneworld on the evolution of the
universe and found that although the effect of the Gauss-Bonnet
correction term in the late time universe is small, it still plays
an important role in the universe evolution.

In contrast to the previous models (\cite{Kirit,Kof,Cai,Bog}), in
our study we kept the energy momentum conservation law on the
brane as usual and found that the combination of the dark
radiation term and the matter content of the bulk can play the
role of the dark energy on the brane and influence the evolution
of the universe. By suitably choosing parameter space in our
model, we can describe the super-acceleration of our universe with
the behavior of the effective dark energy equation of state in
agreement with observations. In \cite{cai2}, it was argued in a
Gauss-Bonnet brane world with induced gravity that the
Gauss-Bonnet term and the mass parameter in the bulk play a
crucial role in the evolution of the universe. Here in our general
model, we confirmed their argument. It is easy to see from
Eqs.(25)-(27) that the Gauss-Bonnet correction influences the
dynamics of our universe, especially in the early universe with
high energy scale. Phenomenon on the Gauss-Bonnet role has been
disclosed in Figs.2-5. We observed that although the Gauss-Bonnet
effect is not clear at the present moment, it influences the
universe evolution in the past and was more important in the
earlier period.

In this work we just restricted our numerical fitting to limited
observational data. Giving the wide range of cosmological data
available, in the future we expect to further constrain our model
parameter space and test the viability of our model.

\acknowledgments {This work was partially supported by NNSF of
China, Ministry of Education of China and Shanghai Educational
Commission and also by Shiraz University.}

\end{document}